\documentclass[twocolumn,showpacs,prb,superscriptaddress]{revtex4}

\usepackage{graphicx}
\usepackage{amssymb}
\usepackage{amsmath}

\begin{document}

\title{Universality in three-dimensional Ising spin glasses: A Monte Carlo 
study}

\author{Helmut G.~Katzgraber}
\author{Mathias K\"orner}
\affiliation
{Theoretische Physik, ETH Z\"urich,
CH-8093 Z\"urich, Switzerland}

\author{A.~P.~Young}
\affiliation{Department of Physics,
University of California,
Santa Cruz, California 95064, USA}

\date{\today}

\begin{abstract}
We study universality in three-dimensional Ising spin glasses by
large-scale Monte Carlo simulations of the Edwards-Anderson Ising
spin glass for several choices of bond distributions, with particular 
emphasis on Gaussian and bimodal interactions. A finite-size 
scaling analysis suggests that three-dimensional Ising spin glasses obey 
universality.
\end{abstract}

\maketitle

\section{Introduction}
\label{sec:introduction}

One of the cornerstones of the theory of critical phenomena is the concept of
universality, according to which the values of many
quantities, such as critical exponents, do not depend on microscopic details
but only on a few broad features such as the dimensionality of space and the
symmetry of the order parameter. Universality follows from renormalization 
group (RG) theory, according to which many interactions that could be 
added to the Hamiltonian are ``irrelevant,'' i.e., do not change the 
critical behavior. The ``$\epsilon$-expansion'' implementation of RG has
been very successful in predicting which perturbations 
are actually relevant and which are irrelevant.
At least for pure systems, and disordered systems without frustration,
numerical simulations seem to be consistent with universality. 

However, for systems with both frustration and disorder, known as spin
glasses,\cite{binder:86} the situation is less clear. On the one hand,
$\epsilon$-expansion calculations as well as high-temperature series 
expansions\cite{daboul:04} for spin glasses imply universality, and,
in the opinion of the authors of this paper, there is no \textit{a priori}
reason why universality should be less valid for spin glasses than for pure 
systems. However, as we shall see below, numerical results so far have not 
been compelling in favor of universality and some
groups\cite{bernardi:96,mari:99,mari:01,henkel:05,pleimling:05} even claim 
explicitly that universality \textit{is violated}.\cite{pleim} Unfortunately,
it is difficult to obtain accurate 
critical exponents because
there are significant corrections to scaling, there are long
equilibration times in Monte Carlo simulations that limit the available system
sizes, and all quantities need to be averaged over many realizations of the
disorder in order to have small enough error bars.

According to universality, the range of the interactions is irrelevant, as
long as it is finite, and so, for example, adding next-nearest-neighbor 
couplings to a nearest-neighbor model will not change the critical behavior. 
However, random systems are characterized not just by the strength of 
first-neighbor, second-neighbor, etc.,~interactions, but by the 
\textit{distributions} of these (random) quantities. Hence, even if one 
restricts oneself to nearest-neighbor interactions, there are many different 
models characterized by different distributions which are expected to be
in the same universality class.  In this paper we attempt to answer, through 
careful simulations, whether this expectation is true for Ising spin
glasses in three dimensions.

\begin{figure}[!tbp]
\includegraphics[width={\columnwidth}]{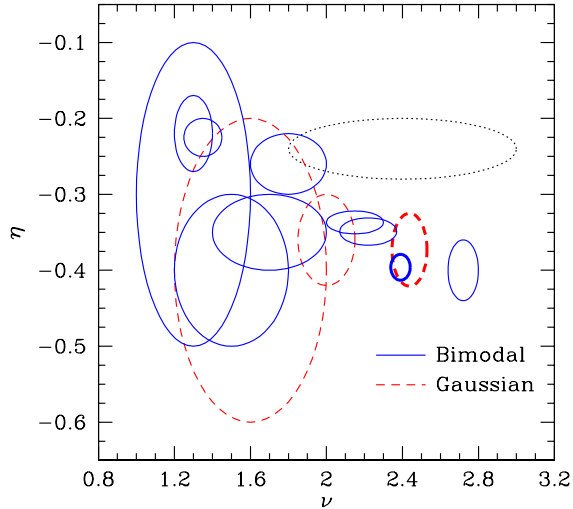}
\vspace*{-1.5cm}
                                                                                
\caption{(Color online)
Graphical representation of different estimates of the critical exponents
$\eta$ and $\nu$ (taken from Table \ref{tab:previous}). The centers of the
ellipses represent the different estimates, and the major axes represent the
error bars of the estimates. The data show a considerable spread. The thick 
lines represent the estimates from the current study. Note that the data 
from the current study for bimodal (solid lines) and Gaussian (dashed lines)
distributions agree within error bars thus providing evidence 
for universality. The dotted line represents data for a random-anisotropy
Heisenberg model in the strong-anisotropy limit (Ref.~\onlinecite{toldin:06}) 
which is expected to be in the same universality class as the Ising spin glass.
}
\label{fig:estimates}
\end{figure}

Many groups have estimated critical exponents for spin glasses for the
Edwards-Anderson (EA) Ising spin glass\cite{edwards:75} in three dimensions 
for different disorder distributions, mainly the Gaussian and bimodal
($\pm J$) models. In Table \ref{tab:previous} we present a summary of 
these results. We also present results from a recent study\cite{joerg:06}
which uses a three-dimensional diluted Ising spin glass
(45\% bond occupation). The advantage of such
a model is that, due to the dilution, cluster algorithms\cite{houdayer:01} 
can be used to study larger system sizes and that corrections to scaling 
seem to be small\cite{joerg:06}
when the bonds are drawn from a bimodal distribution.
The data in Table \ref{tab:previous} show clearly that there is a large 
spread in the estimates of the different critical exponents obtained 
using several different methods, such as series expansions, nonequilibrium
relaxation approaches, and finite-temperature Monte Carlo methods combined
with a finite-size scaling (FSS) analysis. The spread in the different 
estimates of the critical exponents $\nu$ and $\eta$ is easily visualized in
Fig.~\ref{fig:estimates} where we plot $\eta$ versus $\nu$.
As mentioned above, some groups claim universality is 
violated,\cite{bernardi:96,mari:99,mari:01,henkel:05,pleimling:05} but 
most of the papers do not make this claim, probably because the error bars on
the data are usually large. In this paper we aim to test universality 
in equilibrium more precisely by reducing the error bars.

\begin{table}
\caption{
Selection of different estimates (chronologically sorted) of the critical 
temperature $T_{\rm c}$ 
as well as the critical exponents computed by different groups for 
Gaussian (G) as well as bimodal ($\pm J$) random bonds. The estimates
show strong variations and often do not agree. Note that $T_{\rm c}$ 
is \textit{not} universal, so the issue at hand is whether or not the 
results for $\nu$ and for $\eta$ agree 
within the error bars. The results by J\"org (Ref.~\onlinecite{joerg:06}) 
are for a 
bond-diluted $\pm J$ spin glass with 45\% bond occupation, which is why 
the estimate of $T_{\rm c}$ is different than for the standard bimodal 
spin glass. The results of Toldin {\em et al.} (Ref.~\onlinecite{toldin:06}) 
are for 
a random-anisotropy Heisenberg model (RA) in the strong-anisotropy limit,
which is expected to be in the same universality class as the Ising
spin glass in three dimensions.
The last two rows describe results of the present 
study and will be described in detail in what follows.
\label{tab:previous}
}
\begin{tabular*}{\columnwidth}{@{\extracolsep{\fill}} l l l l l }
\hline
\hline
Authors		                              &          & $T_{\rm c}$  & $\nu$     & $\eta$      \\
\hline
Ogielski \& Morgenstern\cite{ogielski:85a}    & $\pm J$  & 1.20(5)      & 1.2(1)    &             \\
Ogielski\cite{ogielski:85}	 	      & $\pm J$  & 1.175(25)    & 1.3(1)    & -0.22(5)    \\
McMillan\cite{mcmillan:85}                    & G        & 1.0(2)       & 1.8(5)    &             \\
Singh \& Chakravarty\cite{singh:86}	      & $\pm J$  & 1.2(1)       & 1.3(2)    &             \\
Bray \& Moore\cite{bray:85}                   & G        & 1.2(1)       & 3.3(6)    &             \\
Bhatt \& Young\cite{bhatt:85}                 & $\pm J$  & 1.2(2)       & 1.3(3)    & -0.3(2)     \\
Bhatt \& Young\cite{bhatt:88}                 & G        & 0.95(5)      & 1.6(4)    & -0.4(2)     \\
Kawashima \& Young\cite{kawashima:96}	      & $\pm J$  & 1.11(4)      & 1.7(3)    & -0.35(5)    \\
Bernardi {\em et al.}\cite{bernardi:96}       & $\pm J$  & 1.165(10)    &           & -0.245(20)  \\
                                              & G        & 0.88(5)      &           & -0.50(4)    \\
I\~nigues {\em et al.}\cite{inigues:96}       & G        & 1.02(5)      & 1.5(3)    &             \\
Berg \& Janke\cite{berg:98}		      & $\pm J$  & 1.12(1)      &           & -0.37(4)    \\ 
Marinari {\em et al.}\cite{marinari:98}       & G        & 0.95(4)      & 2.00(15)  & -0.36(6)    \\
Palassini \& Caracciolo\cite{palassini:99b}   & $\pm J$  & 1.156(15)    & 1.8(2)    & -0.26(4)    \\
Mari \& Campbell\cite{mari:99}                & $\pm J$  & 1.20(1)      &           & -0.21(2)    \\
                                              & G        & 0.86(2)      &           & -0.51(2)    \\
Ballesteros {\em et al.}\cite{ballesteros:00} & $\pm J$  & 1.138(10)    & 2.15(15)  & -0.337(15)  \\
Mari \& Campbell\cite{mari:01}                & $\pm J$  & 1.190(15)    &           & -0.20(2)    \\
                                              & G        & 0.920(15)    &           & -0.42(2)    \\
Mari \& Campbell\cite{mari:02}                & $\pm J$  & 1.195(15)    & 1.35(10)  & -0.225(25)  \\
Nakamura {\em et al.}\cite{nakamura:03}       & $\pm J$  & 1.17(4)      & 1.5(3)    & -0.4(1)     \\
Pleimling \& Campbell\cite{pleimling:05}      & $\pm J$  & 1.19(1)      &           & -0.22(2)    \\
                                              & G        & 0.92(1)      &           & -0.42(2)    \\
J\"org\cite{joerg:06}                         & $\pm J$  & 0.663(6)     & 2.22(15)  & -0.349(18)  \\
Campbell {\em et al.}\cite{campbell:06}       & $\pm J$  &              & 2.72(8)   & -0.40(4)    \\
Toldin {\em et al.}\cite{toldin:06}	      & RA       & 0.93(4)      & 2.4(6)    & -0.24(4)    \\
\hline
This study                                    & G        & 0.951(9)     & 2.44(9) & -0.37(5)      \\
					      & $\pm J$  & 1.120(4)      & 2.39(5) & -0.395(17)   \\

\hline
\hline
\end{tabular*}
\end{table}

A major problem with reducing error bars in critical exponents is the presence
of corrections to FSS, which means that the scaling expressions
used to determine exponents do not work well for small system sizes.
For pure systems, several methods have been proposed in an attempt to reduce
errors caused by corrections to scaling.

First, try to eliminate the leading correction.
In this approach, the model is
altered until the operator which gives the leading correction does not appear
in the Hamiltonian. This means that its effects will not be felt in
\textit{any} calculated quantity. For the three-dimensional Ising ferromagnet,
using a ``soft-spin'' model rather than ``hard'' $\pm 1$ spins, and varying
the coefficient of the fourth-order term in the Hamiltonian, it was possible to
eliminate the leading correction to FSS and obtain high-precision values for
the critical exponents.\cite{hasenbusch:99} We have attempted to do this for 
the spin glass by (i) choosing different disorder distributions, and (ii) 
using soft spins as well as hard spins. However, the corrections to scaling for 
small sizes always had the same sign. Consequently, we have not found a model 
where we could set the leading correction to zero by fine tuning a parameter
in the Hamiltonian. We are not claiming that such a model does not exist; 
only that we were not able to find it.

Second, include corrections to FSS in the analysis. Correction terms
are characterized by an exponent, which is universal, and an amplitude 
which is not. Corrections
to scaling in a scaling plot, where one attempts to collapse 
data from different sizes onto a single curve, occur for both the horizontal 
and vertical axes, see e.g.,~Ref.~\onlinecite{beach:05}.
Thus, a large number of additional parameters have to 
be determined from the data when corrections are included. For ferromagnets, 
where extremely precise data can be obtained for a very large range of sizes,
this is possible.\cite{beach:05} However, for spin glasses, the range of 
system sizes is more limited because of slow dynamics, even though we have 
used state-of-the-art algorithms and considerable computer time, and the 
statistics are not as good because there are large variations between different 
realizations of the disorder. Hence our attempted fits which included 
corrections to scaling did not determine the parameters well and frequently 
the nominal ``best'' fit had extremely large corrections to scaling
and unphysical values for the parameters.

Since attempts to eliminate (leading) corrections to scaling and to explicitly
incorporate them failed, we resorted to the strategy of just including data
for the
larger sizes where we expect corrections to be small, and
neglecting corrections to scaling.
Our main conclusion is
that in equilibrium \textit{universality is satisfied}, since results for a 
particular
observable do not appear to depend on the disorder distribution. However, 
there are still some open questions since \textit{different} observables 
for a \textit{single} disorder distribution yield estimates of critical 
exponents which differ by more than the estimated (statistical) error bars. This 
discrepancy can probably only be resolved by an analysis incorporating 
corrections to scaling. Although this does not appear to be possible at 
present (since the range of sizes is too small) it may be possible in the 
future if more significantly efficient algorithms can be developed. 

The paper is structured as follows: In Sec.~\ref{sec:model} we introduce the
model as well as the measured observables and in Sec.~\ref{sec:numerics} we
present details on the Monte Carlo method used. In Sec.~\ref{sec:analysis} we
discuss the method used to estimate unbiased error bars for the different
estimates of the critical parameters. Results are presented in
Sec.~\ref{sec:results}, followed by concluding remarks in
Sec.~\ref{sec:conclusions}.

\section{Model and observables}
\label{sec:model}
\subsection{Edwards-Anderson Model}

The Hamiltonian of the Edwards-Anderson Ising spin 
glass\cite{edwards:75,binder:86} is given by
\begin{equation}
{\mathcal H} = -\sum_{\langle i,j \rangle} J_{ij} S_i S_j ,
\label{eq:ham}
\end{equation}
where the sites $i$ lie on a three-dimensional cubic lattice of 
size $N = L^3$ and the spins $S_i$ can take values $\pm 1$. The sum is over 
nearest-neighbor pairs and the interactions $J_{ij}$ are independent random 
variables. Periodic boundary conditions are applied.
In this work we mainly study two paradigmatic cases of the EA 
model: 
\begin{itemize}
\item
Gaussian-distributed random bonds with zero mean and standard deviation unity.
\begin{equation}
{\mathcal P}(J_{ij}) = \frac{1}{\sqrt{2\pi}}\, e^{-J_{ij}^2/2} \, .
\label{eq:model_gauss}
\end{equation}
\item
Bimodal ($\pm J$) distribution of bonds in which $J_{ij}$ take the values 
$\pm 1$ with equal probability.
\begin{equation}
{\mathcal P}(J_{ij}) = \frac{1}{2} \left[ \delta (J_{ij} - 1) + 
\delta(J_{ij} + 1)
\right] \, .
\label{eq:model_pmj}
\end{equation}
\end{itemize}
In all cases that we study the mean of the distribution is zero.
Since we set the standard deviation to be unity, the temperature is a
dimensionless quantity.

\subsection{Other Models}
We have also studied other models, although in less detail than the Gaussian 
and $\pm J$ models, in order to see if, by tuning parameters, we could 
eliminate the leading correction to scaling in any of them. These attempts 
have been unsuccessful; therefore we have collected less good statistics for
these models than for the Gaussian and $\pm J$ distributions. Hence we shall
not present results for these other models in
detail, except in Sec.~\ref{sec:global} where we will do a global comparison 
of \textit{all} the models studied to test for universality. These other 
models are 
\begin{itemize}
\item
Gaussian/bimodal distribution with $\sum_{\langle i,j\rangle} J_{ij} = 0$:
This model is the same as the model presented in Eqs.~(\ref{eq:model_gauss}) 
and (\ref{eq:model_pmj}), but with the constraint that the sum of the 
$J_{ij}$ is exactly zero.

\item
Correlated bonds: In this model the nearest-neighbor bonds have
correlations. The probability distribution for the bonds is taken to be
\begin{equation}
{\;\;\;\;\;\;\;} {\mathcal P}\left(J_{ij}\right) \propto \exp \! \left[ 
- \frac{1}{2} \! \sum_{\langle i,j\rangle} J_{ij}^2 -
\lambda \! \sum_{\Box,i} \!\! J_{i1} J_{i2} J_{i3} J_{i4} \right], 
\label{eq:bondham}
\end{equation}
where the last term involves the product of the four bonds around an 
elementary plaquette of the lattice, and is summed over all plaquettes. 
It is this term which generates correlations in the bonds. We
take 
$J_{ij} = \pm 1$, for which the first term in Eq.~(\ref{eq:bondham}) 
is actually a constant. We generate correlated
bonds by first
performing a Monte Carlo simulation \textit{on the bonds} using the 
Hamiltonian in Eq.~(\ref{eq:bondham}). The bonds are then frozen and the spin
simulation is carried out.

\item
Cosine disorder distribution: The bonds of the EA spin glass are chosen
according to 
\begin{equation}
{\mathcal P}(J_{ij}) = 
\frac{1}{2} \, \cos J_{ij} \;\;\;\;
\left[-\frac{\pi}{2} < J_{ij} < \frac{\pi}{2}\right].
\end{equation}

\item
Soft spins: In this model the spins $S_i$ can take any length from 
$-\infty$ to $+\infty$ and the Hamiltonian is given by\cite{hasenbusch:99}
\begin{equation}
{\;\;\;\;\;\;}{\mathcal H} = 
-\sum_{\langle i,j \rangle} J_{ij} S_i S_j + \sum_i S_i^2 +
\lambda \sum_i \left(S_i^2 - 1\right)^2\!\!,
\end{equation}
where the bonds $J_{ij}$ are Gaussian distributed and the parameter $\lambda$
controls the average length of the spins. For $\lambda \to \infty$ we recover
the Ising model with fixed-length spins.

\end{itemize}

\subsection{Measured Quantities}
\label{sec:measured}
We measure different
observables which, in the past, have proven to show a good signature of the
phase transition. First, we study the Binder cumulant\cite{binder:81} given by
\begin{equation}
g = \frac{1}{2}
\left(
3 - \frac{[\langle q^4\rangle_T]_{\rm av}}{[\langle q^2\rangle_T]_{\rm av}^2}
\right) \; ,
\label{eq:binder}
\end{equation}
where $\langle \cdots \rangle_T$ represents a thermal average, 
$[\cdots]_{\rm av}$ is a disorder average, and $q$ is the spin overlap given by
\begin{equation}
q = \frac{1}{N}\sum_{i = 1}^N S_i^\alpha S_i^\beta \; .
\label{eq:q}
\end{equation}	
In the previous equation ``$\alpha$'' and ``$\beta$'' represent two copies of
the system with the same disorder. The Binder ratio is dimensionless and thus
has the simple scaling form
\begin{equation}
g = \widetilde{G}\left(A L^{1/\nu}[\beta - \beta_{\rm c}]
\right)  \; ,
\label{eq:g_scale}
\end{equation}
where $\beta = 1/T$ and $\beta_{\rm c}$ 
is the inverse of the critical temperature.
In addition to $\beta_{\rm c}$, the ``metric factor''\cite{privman:84}
$A$ is also nonuniversal, but, since $A$ is included explicitly, the
resulting scaling function $\widetilde{G}(x)$ \textit{is
universal}.\cite{privman:84} For the models studied here with no lattice
anisotropy,\cite{chen:04,chen:05,selke:05} a 
universality class for finite-size scaling functions is
specified by (i)
the \textit{bulk} universality class, (ii) the boundary conditions, and (iii)
the sample shape.
In this work we always use the same boundary conditions (periodic) and the same
sample shape (cubic), so we expect the
same function $\widetilde{G}(x)$ for models which lie in
the same bulk universality class.
According to Eq.~(\ref{eq:g_scale}), data for
$g$ for different sizes should intersect at $T_{\rm c}$. 
Furthermore, the value of
$g$ at the intersection point, which is given by $\widetilde{G}(0)$ is
also universal since the whole function $\widetilde{G}(x)$ is universal.

In addition, we study the finite-size correlation length
$\xi_L$\cite{cooper:82,palassini:99b,ballesteros:00,martin:02} defined by
\begin{equation}
\xi_L = \frac{1}{2 \sin (k_\mathrm{min}/2)}
\left[\frac{\chi_{\rm SG}(0)}{\chi_{\rm SG}({\bf k}_\mathrm{min})} 
- 1\right]^{1/2},
\label{eq:xiL}
\end{equation}
where ${\bf k}_\mathrm{min} = (2\pi/L, 0, 0)$ is the smallest nonzero
wave vector. Here, the wave vector dependent spin-glass susceptibility is 
given by 
\begin{equation}
\chi_{\rm SG}({\bf k}) = \frac{1}{N} \sum_{i, j} 
\left[ \langle S_i S_j \rangle_T^2 \right]_{\rm av}\!
e^{i{\bf k}\cdot({\bf R}_i - {\bf R}_j)} \; .
\label{eq:chisg}
\end{equation}
Like the Binder ratio, 
the finite-size correlation length divided by the system size is a
dimensionless quantity and so scales as
\begin{equation}
\frac{\xi_L}{L} = \widetilde{X}\left(A L^{1/\nu}[\beta - \beta_{\rm c}]
\right) \; ,
\label{eq:ci_scale}
\end{equation}
in which the metric factor $A$ is the \textit{same}\cite{privman:84}
as in Eq.~(\ref{eq:g_scale}). The reason
the metric factors are the same is as follows: By hypothesis,
the argument of \textit{all} FSS functions 
is really $L/ \xi_\infty$, with \textit{no}
metric factor, where  $\xi_\infty$
is the bulk correlation length.  In this form one has separate FSS functions 
for each side of the transition because $1/\xi_\infty = B_{\pm}|\beta -
\beta_{\rm c}|^\nu$ vanishes in a singular manner at criticality. For example,
\begin{eqnarray}
\frac{\xi_L}{L} & = &
\widehat{X}_\pm\left( L / \xi_\infty \right) 
\label{eq:ci_scale2} \\
& = &
\widehat{X}_\pm\left( B_\pm L[\beta - \beta_{\rm c}]^\nu
\right) \\
& = &
\overline{X}_\pm\left(B_\pm^{1/\nu} L^{1/\nu}[\beta - \beta_{\rm c}]
\right) \, ,
\label{eq:ci_scale3}
\end{eqnarray}
in which $\widehat{X}_\pm$ and $\overline{X}_\pm$ are universal.
Comparing Eq.~(\ref{eq:ci_scale3})
with Eq.~(\ref{eq:ci_scale}), we see that
there are universal, i.e.,~\textit{distribution
independent}, factors $c_\pm$ such that
\begin{equation}
A = c_+ (B_+)^{1/\nu} = c_- (B_-)^{1/\nu} \, ,
\label{A}
\end{equation}
and therefore $\widetilde{X}$ and $\overline{X}$ are essentially
the same functions, in the sense that 
\begin{equation}
\widetilde{X}(u) = \left\{
\begin{array}{ll}
\overline{X}_+(u/c_+) & \;\;\;\;\;\;(u > 0), \\
\overline{X}_-(u/c_-) & \;\;\;\;\;\;(u < 0). 
\end{array}
\right.
\end{equation}
If we repeat the argument for the Binder ratio $g$, we obtain
\begin{equation}
g = \overline{G}_\pm\left(B_\pm^{1/\nu} L^{1/\nu}[\beta - \beta_{\rm c}]
\right) \, ,
\end{equation}
and choosing the same $c_\pm$ as in Eq.~(\ref{A}) we 
reproduce Eq.~(\ref{eq:g_scale}) with the \textit{same} value for $A$ as in
Eq.~(\ref{eq:ci_scale}).

The advantage of $\xi_L/L$ over the Binder ratio $g$ is that the Binder
ratio, being restricted to the interval $g \in [0,1]$, does not have much room
to ``splay out'' below $T_{\rm c}$. 
Presumably because of this, the data for $g$ in
three dimensions depend only very weakly on the system size in this 
region.\cite{kawashima:96,marinari:98}
However, the finite-size correlation length $\xi_L/L$
does not have this constraint, and so the data for it
splays out better at low temperatures, allowing for a more precise
determination of the critical parameters.

Both $g$ and $\xi_L/L$ allow one to determine $T_{\rm c}$ and the critical 
exponent $\nu$. However, to fully characterize the critical behavior of a 
system, a second critical exponent, $\eta$, is 
required.\cite{yeomans:92} Thus we also study the scaling of the spin-glass 
susceptibility $\chi_{\rm SG} \equiv \chi_{\rm SG}({\bf k} = 0)$ given by 
Eq.~(\ref{eq:chisg}) with ${\bf k} = 0$. Near criticality we expect
\begin{equation}
\chi_{\rm SG} = D L^{2 - \eta}
\widetilde{C}\left(A L^{1/\nu}[\beta - \beta_{\rm c}]\right) 
\label{eq:chisg_scale}
\end{equation}
therefore allowing us to determine the critical exponent $\eta$. By separating
out the nonuniversal amplitude $D$, the scaling function $\widetilde{C}$
is universal.

\section{Numerical details}
\label{sec:numerics}

The simulations are done using the parallel tempering Monte Carlo
method.\cite{hukushima:96,marinari:96} For the Gaussian distribution we 
have tested equilibration with the method introduced in 
Ref.~\onlinecite{katzgraber:01} where the energy computed directly is
compared to the energy computed from the link overlap. The data for both
quantities approach a limiting value from opposite directions. 
Once they agree, and other
observables are independent of Monte Carlo steps, the system is in
equilibrium. For the bimodal disorder distribution
we use a multispin coded version of the algorithm which allows us to update
32 copies of the system at the same time.
The aforementioned equilibration test cannot be applied to the
bimodal spin glass. In this case we study how the results vary when the
simulation time is successively increased by factors of 2 (logarithmic
binning). We require that the last three results for all observables 
agree within error bars.

Parameters of the simulation are presented in Tables \ref{tab:simparams_g} and
\ref{tab:simparams_pmj} for the Gaussian and bimodal ($\pm J$) distributions,
respectively.

\begin{table}
\caption{
Parameters of the simulations for Gaussian distributed disorder. 
$N_{\rm sa}$ is the number of samples, $N_{\rm sw}$ is the total number 
of Monte Carlo sweeps for each of the $2 N_T$ replicas for a single sample,
$T_{\rm min}$ is the lowest temperature simulated, and $N_T$ is the number
of temperatures used in the parallel tempering method for each system size
$L$.
\label{tab:simparams_g}
}
\begin{tabular*}{\columnwidth}{@{\extracolsep{\fill}} c r r r r l }
\hline
\hline
$L$  &  $N_{\rm sa}$  & $N_{\rm sw}$ & $T_{\rm min}$ & $N_{T}$  \\
\hline
 $3$ &  $20000$ &   $32768$ & $0.80$ &  $8$ \\
 $4$ &  $20000$ &   $20000$ & $0.80$ &  $8$ \\
 $6$ &  $20000$ &   $40000$ & $0.80$ &  $8$ \\
 $8$ &  $20000$ &   $50000$ & $0.80$ & $10$ \\
$12$ &  $10000$ &  $655360$ & $0.80$ & $16$ \\
$16$ &  $ 5000$ & $1048576$ & $0.80$ & $33$ \\
\hline
\hline
\end{tabular*}
\end{table}

\begin{table}
\caption{
Parameters of the simulations, defined in Table~\ref{tab:simparams_g},
for bimodal distributed disorder.
The system sizes marked with an asterisk have been simulated with 
the more efficient multispin method.
\label{tab:simparams_pmj}
}
\begin{tabular*}{\columnwidth}{@{\extracolsep{\fill}} c r r r r l }
\hline
\hline
$L$  &  $N_{\rm sa}$  & $N_{\rm sw}$ & $T_{\rm min}$ & $N_{T}$  \\
\hline
 $3$   &  $40000$ &    $8000$ & $0.82$ & $16$ \\
 $4$   &  $40000$ &    $8000$ & $0.82$ & $16$ \\
 $6$   &  $40000$ &   $20000$ & $0.82$ & $16$ \\
 $8$   &  $30000$ &   $80000$ & $0.82$ & $16$ \\
$12$   &  $15807$ &  $300000$ & $0.82$ & $18$ \\
$16^*$ &  $11360$ &  $128000$ & $0.95$ & $16$ \\
$20^*$ &  $ 9408$ & $1280000$ & $1.05$ & $25$ \\
$24^*$ &  $ 8416$ & $1280000$ & $1.05$ & $25$ \\
\hline
\hline
\end{tabular*}
\end{table}

\section{Statistical Analysis of the Data}
\label{sec:analysis}

For the Binder ratio and finite-size correlation length we need to find the 
best choice of parameters, $\nu$ and $\beta_{\rm c}$ 
in order to collapse the data 
onto the scaling predictions of Eqs.~(\ref{eq:g_scale}) and 
(\ref{eq:ci_scale}), respectively. 
To do so we assume that the scaling function can be
represented by a third-order polynomial $y(x) = c_0 + c_1x + c_2x^2 + c_3x^3$
and do a global fit to the six 
parameters $c_i, (i=0, \cdots, 3), \beta_{\rm c}$, 
and $\nu$. We also analyze results for $\chi_{\rm SG}$, for which
there is a seventh parameter, the critical
exponent $\eta$. We have performed these nonlinear fits in two ways: (i) 
a code based on the Levenberg-Marquardt
algorithm,\cite{press:95} and (ii) the statistics package R.\cite{R} 
The same results have been obtained from both approaches.

It is also necessary to obtain error bars on the fit parameters.
One has to be careful because, for a given size, all temperatures are
simulated with the same disorder realization in the parallel tempering Monte 
Carlo method.\cite{hukushima:96,marinari:96} Hence the fitted data are
correlated. We therefore have applied the
following procedure: For each system size $L$ with $N_{\rm sa}$ disorder
realizations, a
randomly selected bootstrap\cite{bootsamp}
sample of $N_{\rm sa}$ disorder realizations
is generated. With this random sample, an estimate of the different 
observables (with bootstrap error bars) is computed for each temperature.
We repeat this procedure $N_{\rm boot} = 1000$ times for each lattice size 
and then assemble $N_{\rm boot}$ complete data sets
(each having results for every size) by combining the $i$-th bootstrap
sample 
for each size for $i = 1, \cdots, N_{\rm boot}$. The finite-size scaling fit 
described above is then carried out on each of these $N_{\rm boot}$ sets, 
thus obtaining $N_{\rm boot}$ estimates of the fit parameters.
Since the bootstrap sampling is done with respect to the disorder realizations
which are statistically independent we can use a conventional
bootstrap analysis to estimate statistical error bars on the fit parameters.
These are equal to the standard deviation among the $N_{\rm boot}$ bootstrap 
estimates.

\section{Results}
\label{sec:results}

\subsection{Gaussian-distributed random bonds}
\label{sec:gauss}

\begin{figure}[!tbp]
\includegraphics[width={\columnwidth}]{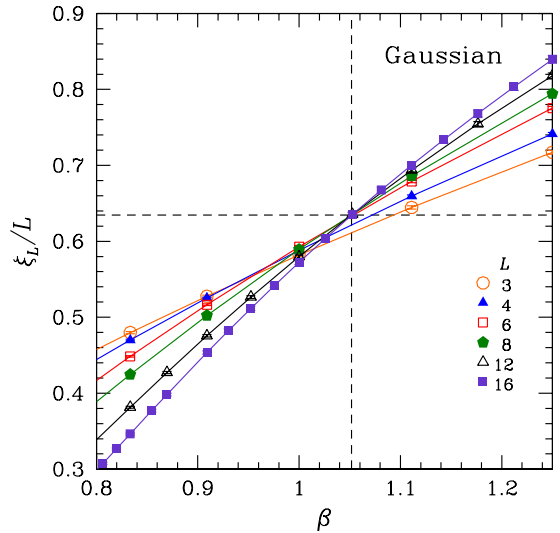}
\vspace*{-1.5cm}

\caption{(Color online)
Finite-size correlation length $\xi_L/L$ as a function of inverse temperature
$\beta$ for the three-dimensional Edwards-Anderson Ising spin glass with
Gaussian disorder for several system sizes $L$. The data cross at 
$\beta_{\rm c}^{-1} \simeq 0.951$. The dashed lines represent the optimal
values obtained from a finite-size scaling for $\beta_{\rm c}$ and
$\xi_L(\beta_{\rm c})/L$.
}
\label{fig:xi_gauss}
\end{figure}

\begin{figure}[!tbp]
\includegraphics[width=\columnwidth]{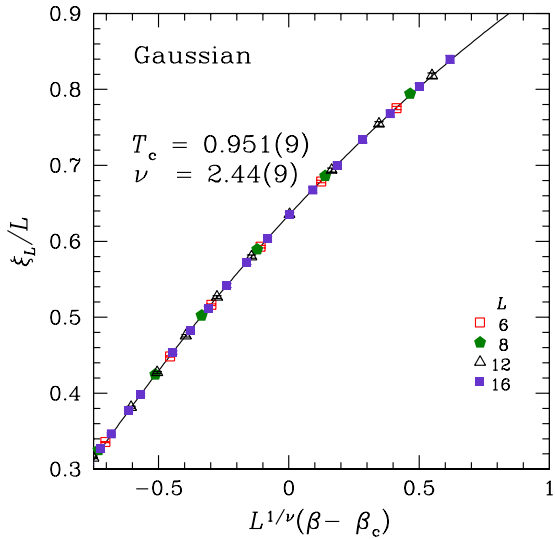}
\vspace*{-1.5cm}

\caption{(Color online)
Finite-size scaling analysis of $\xi_L/L$ for the Ising spin glass with 
Gaussian bonds according to Eq.~(\ref{eq:ci_scale}). The sizes are 
$6 \le L \le 16$. The solid line is the third-order polynomial used in the
fit, see Sec.~\ref{sec:analysis}.
}
\label{fig:xi_scale_gauss}
\end{figure}

\begin{table}[!tbp]
\caption{
Summary of critical parameters for a Gaussian disorder distribution estimated
by scaling the data. Scaling has been done in $(\beta - \beta_{\rm c})$ 
except for the spin-glass susceptibility for which the data has been scaled 
with 
$(T - T_{\rm c})$. In the table below, $T_{\rm c}$ is the critical temperature,
and $\eta$ and $\nu$ are critical exponents. The quantity $c_0$ is the 
zeroth-order coefficient of the fitting polynomial and corresponds to the 
value of a given observable at criticality. $\chi^2$ represents the 
chi-squared value for the finite-size scaling fitting 
function (Ref.~\onlinecite{press:95}). 
For comparison the number of data points used in the 
fit is 25.
For the fit using the scaling form of Ref.~\onlinecite{campbell:06} the value
of $T_{\rm c}$ is fixed to be that obtained from $\xi_L/L$.
\label{tab:critparams_gauss}
}
\begin{tabular*}{\columnwidth}{@{\extracolsep{\fill}} l l l l  }
\hline
\hline
$\xi_L/L$            & Estimate     & Error         \\
\hline
$c_0$                & 0.6346       & 0.0090        \\
$T_{\rm c}$          & 0.9508       & 0.0089        \\
$\nu$                & 2.4370       & 0.0924        \\
$\chi^2$             & 11.7859      & 10.2696       \\
\hline
\hline
$g$                  & Estimate     & Error         \\
\hline
$c_0$                & 0.7600       & 0.0068        \\
$T_{\rm c}$          & 0.9310       & 0.0137        \\
$\nu$                & 2.6761       & 0.1662        \\
$\chi^2$             & 17.7245      & 14.0111       \\
\hline
\hline
$\chi_{\rm SG}$	     & Estimate     & Error    \\
\hline
$T_{\rm c}$          & 0.9489       & 0.0264        \\
$\nu$                & 1.4859       & 0.0602        \\
$\eta$               & -0.3733      & 0.0483        \\
$\chi^2$             & 12.8776      & 8.0025        \\
\hline
\hline
$\chi_{\rm SG}$ (scaling as in Ref.~\onlinecite{campbell:06})& Estimate  & Error   \\
\hline
$T_{\rm c}$          & 0.9508       & 0.0089   \\
$\nu$                & 2.7767       & 0.0249   \\
$\eta$               & -0.3716      & 0.0055   \\
$\chi^2$             & 18.3403      & 12.6776   \\
\hline
\hline
\end{tabular*}
\end{table}

Figure \ref{fig:xi_gauss} shows data for the finite-size correlation length as
a function of the inverse temperature for different system sizes.
The data for $L \ge 6$ intersect at (or very close to) a common point whereas
the data for the smallest sizes, $L=3$ and 4 lie consistently too low in this
region.
The fact that sizes $L=3$ and $4$ do not intersect at a common point clearly
indicates that corrections to scaling are significant for these sizes.
We were hoping to find other models where the trend would be the other way
around (i.e.,~where the small-$L$ data are too high) so that by fine tuning of 
parameters we could eliminate this correction to scaling. However, all 
the models studied (see Sec.~\ref{sec:model}) had corrections of the 
same sign as shown in Fig.~\ref{fig:xi_gauss}.

In scaling the data according to Eq.~(\ref{eq:ci_scale}) in the way
discussed in Sec.~\ref{sec:analysis}, we omit sizes $L=3$ and
4 because these data are clearly affected by corrections to scaling, and
Fig.~\ref{fig:xi_scale_gauss} shows the resulting plot for sizes $L \ge 6$.  
The overall fit is very satisfactory and gives the critical parameters shown in
Table \ref{tab:critparams_gauss}.

\begin{figure}[!tbp]
\includegraphics[width=\columnwidth]{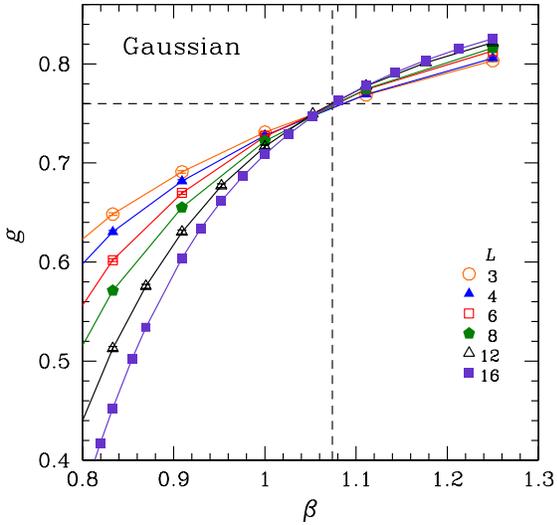}
\vspace*{-1.5cm}

\caption{(Color online)
Binder ratio $g$, defined in Eq.~(\ref{eq:binder}), as a function of 
inverse temperature $\beta$ for the three-dimensional Edwards-Anderson Ising 
spin glass with Gaussian bonds. The data cross at 
$\beta_{\rm c}^{-1} \simeq 0.931$. The dashed lines represent the optimal
values obtained from a finite-size scaling for $\beta_{\rm c}$ and
$g(\beta_{\rm c})$.
}
\label{fig:g_gauss}
\end{figure}

In Fig.~\ref{fig:g_gauss} we show data for the Binder ratio, 
Eq.~(\ref{eq:binder}), as a function of the inverse temperature $\beta$. 
The data cross at $\beta_{\rm c}^{-1} \simeq 0.931$.
Note that for $\beta > \beta_{\rm c}$ the data do not splay very well 
thus making it difficult to determine the critical temperature accurately.

\begin{figure}[!tbp]
\includegraphics[width=\columnwidth]{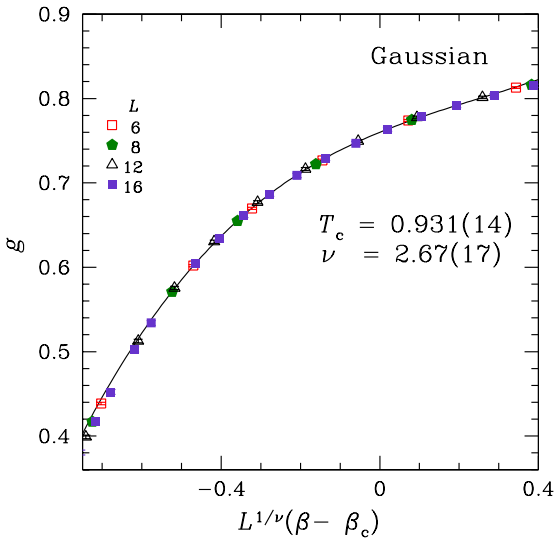}
\vspace*{-1.5cm}

\caption{(Color online)
Finite-size scaling analysis of the data for the Binder ratio $g$
according to Eq.~(\ref{eq:g_scale}) for the three-dimensional Ising spin 
glass with Gaussian disorder. The scaling analysis is performed for $L \ge 6$
and the solid line represents the best fit to the data from the finite-size
scaling analysis.
}
\label{fig:g_scale_gauss}
\end{figure}

\begin{figure}[!tbp]
\includegraphics[width=\columnwidth]{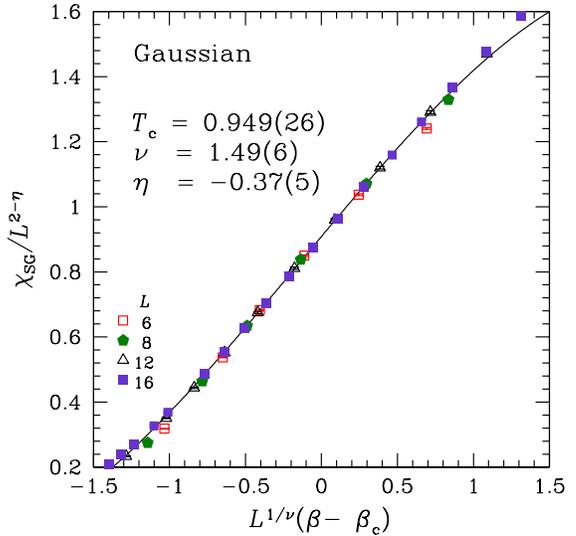}
\vspace*{-1.5cm}

\caption{(Color online)
Finite-size scaling of the spin-glass susceptibility $\chi_{\rm SG}$ 
according to Eq.~(\ref{eq:chisg_scale}) for Gaussian disorder.
In fact, the method of scaling worked
better when using $T$ rather than $\beta$ and the fit parameters shown are for
the $T$ fit. However, we show the resulting scaling plot using $\beta$ for
consistency with the other plots.
The scaling analysis is performed 
for $L \ge 6$.
}
\label{fig:chi_scale_gauss}
\end{figure}

Using the analysis presented in Sec.~\ref{sec:analysis} we have obtained 
the best
fit, shown in Table \ref{tab:critparams_gauss}, and present the scaling plot 
of the Binder ratio in Fig~\ref{fig:g_scale_gauss}. For the analysis we have 
only considered $L \ge 6$. 

Overall, we expect that the analysis of $\xi_L/L$ gives more accurate
results than that for $g$, because the data for $g$ do not splay out much
below $T_{\rm c}$, and so our best results for $T_{\rm c}$ 
and $\nu$ are those for
$\xi_L/L$, i.e.
\begin{equation}
T_{\rm c} = 0.951(9), \quad \nu = 2.44(9) \quad ({\rm Gaussian}) \, .
\label{eq:fitparams_gauss}
\end{equation}
We emphasize that the error bars quoted in this paper
are only statistical. There are also
systematic errors, which are hard to estimate for the range of sizes that can
be studied. We discuss systematic errors below in more detail, especially in
Secs.~\ref{sec:campbell} and \ref{sec:conclusions}.
It is gratifying that
the results obtained from the analysis of
$g$, namely, $T_{\rm c} = 0.931(17)$, $\nu = 2.67(17)$ are (just)
consistent with these. In addition to exponents, the values of $\xi_L/L$ and
$g$ at $T_{\rm c}$ are also expected to be universal. These are given by the
appropriate values of $c_0$ in Table~\ref{tab:critparams_gauss}:
\begin{equation}
\frac{\xi_L(T_{\rm c})}{L} = 0.635(9), \quad 
g(T_{\rm c})= 0.760(70) \ \  ({\rm Gaussian}) \, .
\label{eq:c0_gauss}
\end{equation}

Unfortunately, the situation for the analysis of the spin-glass susceptibility
data is less gratifying.
The scaling plot for the spin-glass susceptibility is shown in
Fig.~\ref{fig:chi_scale_gauss}. For consistency with the other plots the
horizontal axis in this figure is $\beta$. However, the method of fitting
(third-order polynomial) works best, in this case, for a fit using $T$. Hence
Fig.~\ref{fig:chi_scale_gauss} indicates the fit parameters from the $T$ fit.

The results of the fit are
\begin{equation}
T_{\rm c} = 0.949(26), \quad \nu = 1.49(6), \quad 
\eta =  -0.37(5).
\label{eq:eta_gauss}
\end{equation}
The value for $T_{\rm c}$ agrees within the error bars
with those from $\xi_L/L$ and $g$, 
but the value for
$\nu$ is in strong disagreement. Disagreements between exponents obtained
in different ways are presumably due to
corrections to FSS, but the size of the difference here is surprisingly large.
In Sec.~\ref{sec:campbell} we shall revisit the problem of the 
surprisingly low value for $\nu$ obtained from $\chi_{\rm SG}$.

\subsection{Bimodal-distributed random bonds}
\label{sec:pmJ}

\begin{table}[!tbp]
\caption{
Summary of critical parameters for a bimodal disorder distribution estimated
by scaling the data. The scaling is done in $(\beta - \beta_{\rm c})$ except 
for the data for the spin-glass susceptibility where the scaling is done
in $(T - T_{\rm c})$. For further details see the caption of Table
\ref{tab:critparams_gauss}. The number of data points used in the fits is 48.
In the fit for $\chi_{\rm SG}$
using the scaling form of Ref.~\onlinecite{campbell:06} the value
of $T_{\rm c}$ is fixed to be that obtained from $\xi_L/L$.
\label{tab:critparams_pmj}
}
\begin{tabular*}{\columnwidth}{@{\extracolsep{\fill}} l l l l  }
\hline
\hline
$\xi_L/L$            & Estimate     & Error         \\
\hline
$c_0$                & 0.6265       & 0.0036        \\
$T_{\rm c}$          & 1.1199       & 0.0037        \\
$\nu$                & 2.3900       & 0.0514        \\
$\chi^2$             & 52.8369      & 28.3532       \\
\hline
\hline
$g$                  & Estimate     & Error         \\
\hline
$c_0$                & 0.7626       & 0.0029        \\
$T_{\rm c}$          & 1.0881       & 0.0062        \\
$\nu$                & 2.7937       & 0.1103        \\
$\chi^2$             & 60.5020      & 30.2953       \\
\hline
\hline
$\chi_{\rm SG}$      & Estimate     & Error   \\
\hline
$T_{\rm c}$          & 1.1040       & 0.0097        \\
$\nu$                & 1.5721       & 0.0251        \\
$\eta$               & -0.3954      & 0.0168        \\
$\chi^2$             & 88.2526      & 31.8466       \\
\hline
\hline
$\chi_{\rm SG}$ (scaling as in Ref.~\onlinecite{campbell:06})& Estimate  & Error   \\
\hline
$T_{\rm c}$  	     & 1.1199       & 0.0037   \\
$\nu$                & 2.7376       & 0.0166   \\
$\eta$               & -0.3663      & 0.0166   \\
$\chi^2$             & 83.6070      & 36.7894  \\
\hline
\hline
\end{tabular*}
\end{table}

\begin{figure}[!tbp]
\includegraphics[width=\columnwidth]{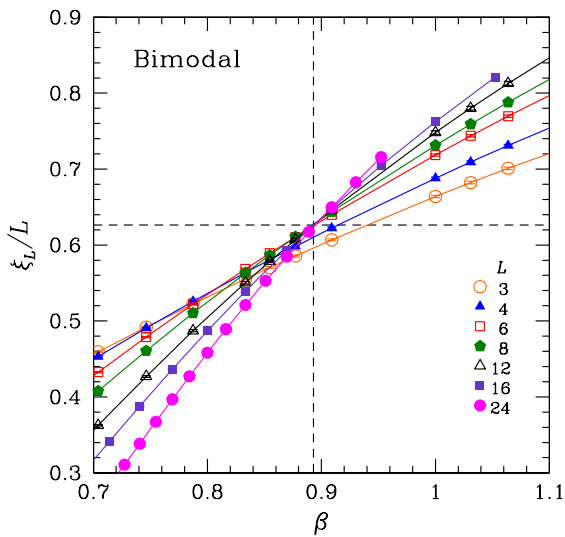}
\vspace*{-1.5cm}

\caption{(Color online)
Finite-size correlation length $\xi_L/L$ as a function of inverse temperature
$\beta$ for the three-dimensional Edwards-Anderson Ising spin glass with
$\pm J$ bonds for several system sizes $L$. The data cross at 
$\beta_{\rm c}^{-1} \simeq 1.12$. The dashed lines represent the optimal
values obtained from a finite-size scaling for $\beta_{\rm c}$ and
$\xi_L(\beta_{\rm c})/L$. 
}
\label{fig:xi_pmJ}
\end{figure}

\begin{figure}[!tbp]
\includegraphics[width=\columnwidth]{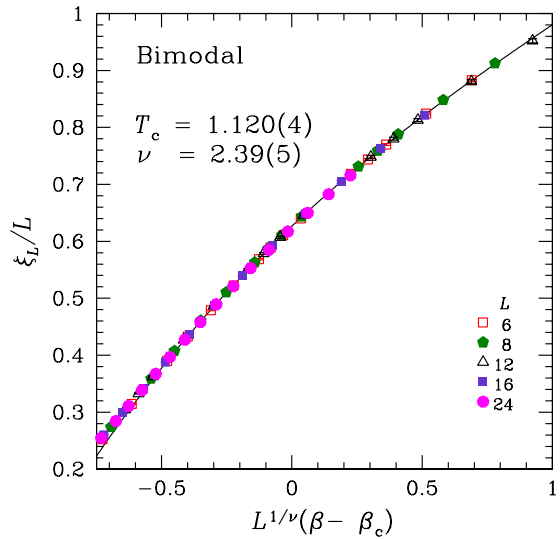}
\vspace*{-1.5cm}

\caption{(Color online)
Finite-size scaling analysis of $\xi_L/L$ for the Ising spin glass 
with $\pm J$ bonds according to Eq.~(\ref{eq:ci_scale}). The scaling analysis 
is performed for $L \ge 6$.
}
\label{fig:xi_scale_pmJ}
\end{figure}

We show data for $\xi_L/L$ in Fig.~\ref{fig:xi_pmJ}. It is fairly similar to
the data for the Gaussian distribution in that the larger sizes show a common
intersection, but the data for the smaller sizes, $L=3$ and $4$,
are lower, showing that these sizes are affected by corrections to FSS. Hence
we only use sizes $6 \le L \le 24$ in the scaling plot which is shown in
Fig.~\ref{fig:xi_scale_pmJ}. Parameters obtained from the fits are shown in
Table~\ref{tab:critparams_pmj}.

\begin{figure}[!tbp]
\includegraphics[width=\columnwidth]{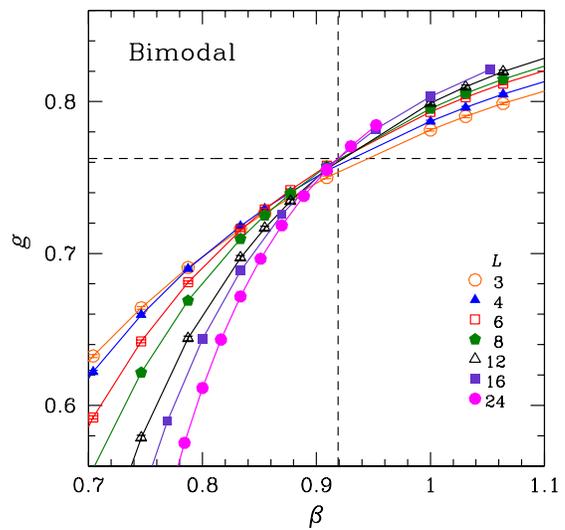}
\vspace*{-1.5cm}

\caption{(Color online)
Binder ratio $g$ as a function of inverse temperature
$\beta$ for the three-dimensional Edwards-Anderson Ising spin glass with
$\pm J$ bonds for several system sizes $L$. The data cross at 
$\beta_{\rm c}^{-1} \simeq 1.088$. The dashed lines represent the optimal
values obtained from a finite-size scaling for $\beta_{\rm c}$ and
$g(\beta_{\rm c})/L$.
}
\label{fig:g_pmJ}
\end{figure}

\begin{figure}[!tbp]
\includegraphics[width=\columnwidth]{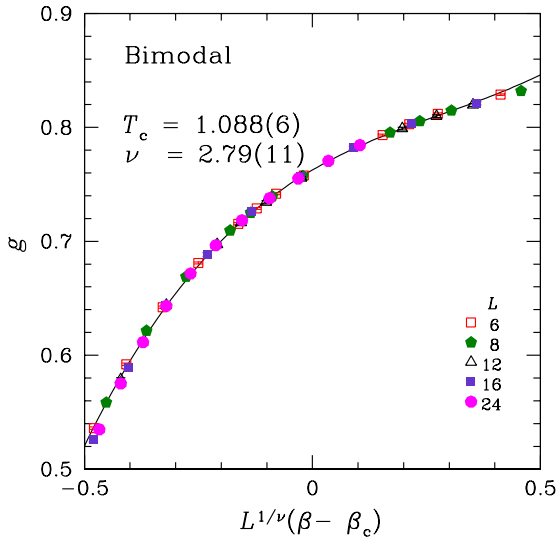}
\vspace*{-1.5cm}

\caption{(Color online)
Finite-size scaling analysis of the data for the Binder ratio $g$
according to Eq.~(\ref{eq:g_scale}) for the three-dimensional Ising spin 
glass with $\pm J$ bonds. The sizes used are $6 \le L \le 24$. 
}
\label{fig:g_scale_pmJ}
\end{figure}

Figure \ref{fig:g_pmJ} shows data for the Binder ratio, Eq.~(\ref{eq:binder}),
for different system sizes $L$ as a function of the inverse temperature
$\beta$. The corresponding scaling plot is shown in
Fig.~\ref{fig:g_scale_pmJ}.

The fit parameters obtained from $\xi_L/L$ are
\begin{equation}
T_{\rm c} = 1.120(4), \quad \nu = 2.39(5) \quad (\pm J) \, .
\label{eq:fitparams_pmJ}
\end{equation}
Those obtained from $g$, namely, $T_{\rm c} = 1.088(6)$, $\nu = 2.79(11)$
disagree somewhat, but, as also
discussed above, we feel that those for $\xi_L/L$ are more reliable because
the data for $\xi_L/L$ splay out more below $T_{\rm c}$.

The values of $\xi_L/L$ and $g$ at $T_{\rm c}$ are given by the appropriate 
values of $c_0$ in Table~\ref{tab:critparams_pmj}:
\begin{equation}
\frac{\xi_L(T_{\rm c})}{L}  =  0.627(4), \quad
g(T_{\rm c})  =  0.763(3) \quad (\pm J) \, .
\label{eq:c0_pmJ}
\end{equation}

\begin{figure}[!tbp]
\includegraphics[width=\columnwidth]{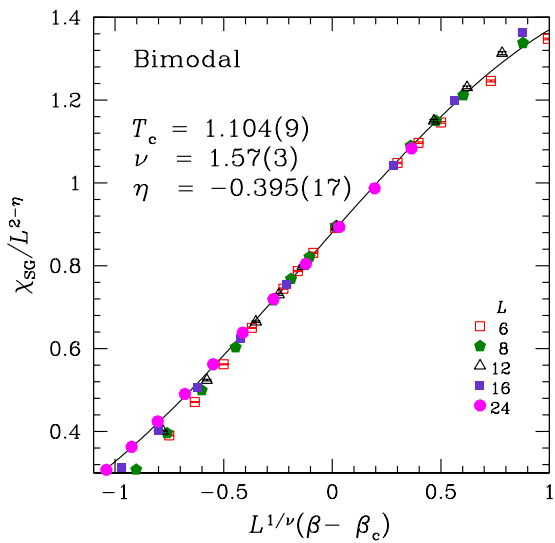}
\vspace*{-1.5cm}

\caption{(Color online)
Finite-size scaling of the spin-glass susceptibility $\chi_{\rm SG}$ according
to Eq.~(\ref{eq:chisg_scale}) for the three-dimensional Ising spin 
glass with $\pm J$ bonds. As for the corresponding plot for the Gaussian
distribution, Fig.~\ref{fig:chi_scale_gauss}, the fit is actually done in 
$T$, but the plot is given as a function of $\beta$ for consistency with 
the other plots.
The scaling analysis is performed for $L \ge 6$. 
}
\label{fig:chi_scale_pmJ}
\end{figure}

A scaling plot for $\chi_{\rm SG}$ is shown in Fig.~\ref{fig:chi_scale_pmJ}.
The best fit (using $T$ rather than $\beta$) gives
\begin{equation}
T_{\rm c} = 1.104(9), \quad \nu = 1.57(3), \quad \eta = -0.395(17).
\label{eq:eta_pmJ}
\end{equation}
The value of $T_{\rm c}$ differs from that obtained from $\xi_L/L$, in
Eq.~(\ref{eq:fitparams_pmJ})
by rather more than the error bars. If we fix $T_{\rm c}$ to be that obtained from
$\xi_L/L$ we find
\begin{equation}
\nu = 1.527(8), \quad \eta = -0.368(24).
\label{eq:eta2_pmJ}
\end{equation}
The value of $\eta$ obtained in this way is consistent with that in the
unconstrained fit in Eq.~(\ref{eq:eta_pmJ}).  The value of $\nu$ is slightly
different from that in Eq.~(\ref{eq:eta_pmJ}), but more importantly, both
these values of $\nu$ obtained from $\chi_{\rm SG}$ are considerably smaller
than those obtained from $\xi_L/L$ and $g$. This is the same situation than
found for the Gaussian distribution.
Interestingly,
the values of $\nu$ from $\chi_{\rm SG}$ for the two distributions
agree quite well \textit{with each other}.

We argue that the most reliable quantity to analyze is $\xi_L/L$ because this
has clean intersections with significant splaying out
below $T_{\rm c}$. The results
for the exponent $\nu$, shown in Eqs.~(\ref{eq:fitparams_gauss}) and
(\ref{eq:fitparams_pmJ}), for the Gaussian and $\pm J$ distributions agree
well within the error bars, which supports universality. Further support for
universality comes from the agreement in the values of $\xi_L/L$ and $g$ at
the critical point, shown in Eqs.~(\ref{eq:c0_gauss}) and
(\ref{eq:c0_pmJ}). We also note the agreement in the values of $\eta$ from
Eqs.~(\ref{eq:eta_gauss}) and (\ref{eq:eta_pmJ}) or (\ref{eq:eta2_pmJ}).

\subsection{Alternative analysis of $\boldsymbol{\chi_{\rm SG}}$}
\label{sec:campbell}

The main unresolved issue is the large difference in the values of $\nu$
obtained from the spin-glass susceptibility compared with those obtained from
$\xi_L/L$ and $g$.
Recently, Campbell {\em et al.}\cite{campbell:06} have claimed that the
difference is much diminished if one uses an alternative scaling form. They
propose that the scaling region will be larger, so one can incorporate data
for a larger range of temperature, if the behavior as $T \to \infty$ is
consistent with the scaling function. To be precise, they propose that
\begin{eqnarray}
g & = & 
\widetilde{G} \left[(L T)^{1/\nu}\left( 1 - (T_{\rm c} / T)^2 \right) \right], 
\label{eq:g_campbell} \\
\frac{\xi_L}{L} & = &
\widetilde{X} \left[(L T)^{1/\nu}\left( 1 - (T_{\rm c} / T)^2 \right) \right], 
\label{eq:xi_campbell} \\
\chi_{\rm SG}  & = & (L T)^{2-\eta} \,
\widetilde{C} \left[(L T)^{1/\nu}\left( 1 - (T_{\rm c} / T)^2 \right) \right] ,
\label{eq:chisg_campbell} 
\end{eqnarray}
where we have not included explicitly the metric factors.
Asymptotically, for $L \to \infty$ and $(T-T_{\rm c}) \to 0$, these expressions are
equivalent to the standard forms that we have used, Eqs.~(\ref{eq:g_scale}),
(\ref{eq:ci_scale}),
and (\ref{eq:chisg_scale}).  Thus the difference between the
expressions proposed by Campbell {\em et al}.~and the standard expressions 
is only in the corrections to scaling.

In both the original scaling forms and the modified form of Campbell
{\em et al.}, $T_{\rm c}$ is located by intersections of data for $\xi_L/L$ and $g$ of
different sizes. Thus we do not expect the estimates of $T_{\rm c}$ to be very
different in the two approaches. Furthermore, if we restrict data to the
region close to $T_{\rm c}$, the data collapse involves mainly the derivative of the
data with respect to temperature at $T_{\rm c}$.
For $\xi_L/L$ and $g$, both the original
and modified form predict that the temperature derivative is proportional to
$L^{1/\nu}$. Hence, for $\xi_L/L$ and $g$,
we also do not expect very different values for $\nu$ from
the two scaling forms. These expectations are confirmed by
our analysis. Using
Eqs.~(\ref{eq:g_campbell}) and (\ref{eq:xi_campbell}), we find values for $T_{\rm c}$
and $\nu$ which agree, within the error bars, with those described above in
Secs.~\ref{sec:gauss} and \ref{sec:pmJ} which used the standard scaling
forms. It is possible that scaling may work for data over a larger range of $(T
- T_{\rm c})$, at least above\cite{asymmetry} $T_{\rm c}$, using
Eqs.~(\ref{eq:g_campbell}) and (\ref{eq:xi_campbell}),
but we have not investigated this in detail.

However, the situation for $\chi_{\rm SG}$ is quite different, because of the
factor of $T^{2-\eta}$ in front of the scaling function in 
Eq.~(\ref{eq:chisg_campbell}), since
\begin{equation}
\left.
\frac{1}{\chi_{\rm SG}} \, \frac{d\, \chi_{\rm SG}}{dT}\right|_{T_{\rm c}}  = 
a L^{1/\nu} + b \,  ,
\label{eq:deriv}
\end{equation}
where $a$ and $b$ depend on $T_{\rm c}$ and the value of the scaling function and 
its derivative at zero argument, but not on $L$.  The factor of $b$ arises
from the $T$ dependence of the prefactor \textit{outside} the scaling
function in Eq.~(\ref{eq:chisg_campbell}),
and does not occur in the analogous expressions for $\xi_L/L$ and
$g$ in Eqs.~(\ref{eq:xi_campbell}) and (\ref{eq:g_campbell}).
For $L \to \infty$, the $L^{1/\nu}$ term in Eq.~(\ref{eq:deriv})
dominates and we recover the same behavior as in the standard scaling form.
However, for the small sizes that can be studied numerically, the factor of
$b$ gives a significant correction to scaling \textit{especially since $1/\nu$
is small}. This is why Campbell {\em et al.}\cite{campbell:06} found a large
difference in the value of $\nu$ obtained from from $\chi_{\rm SG}$ using
their scaling compared with conventional finite-size scaling. 

We also find a large difference. If we fix $T_{\rm c}$ to be the value
obtained from $\xi_L/L$ we obtain, from Eq.~(\ref{eq:chisg_campbell}) 
\begin{eqnarray}
\nu & = & 2.777(25), \ \quad \eta = -0.372(6) \ \quad {\rm (Gaussian)} , 
\label{cam_scal_gauss} \\
\nu & = & 2.738(17), \ \quad \eta = -0.366(3) \ \quad (\pm J), 
\label{cam_scal_pmJ} 
\end{eqnarray}
see also Tables \ref{tab:critparams_gauss} and \ref{tab:critparams_pmj}.
These values for $\nu$
are \textit{considerably} larger than those found from the standard
analysis
discussed in Secs.~\ref{sec:gauss} and \ref{sec:pmJ}.
They are even somewhat larger than those found from $\xi_L/L$,
although they agree better with the $\xi_L/L$ values than those found from
$\chi_{\rm SG}$ using the standard analysis.
The fact that we, like
Ref.~\onlinecite{campbell:06}, obtain very different values for $\nu$ from 
$\chi_{\rm SG}$ depending on the form of the scaling function used, tells us
that corrections to scaling can be very important in spin glasses for the
range of sizes that can be simulated. We note, however, that the two estimates
in Eqs.~(\ref{cam_scal_gauss}) and (\ref{cam_scal_pmJ})
agree well \textit{with each other}, so we still find no evidence for
lack of universality.

\subsection{Global comparisons of all the models}
\label{sec:global}

\begin{figure}
\includegraphics[width=\columnwidth]{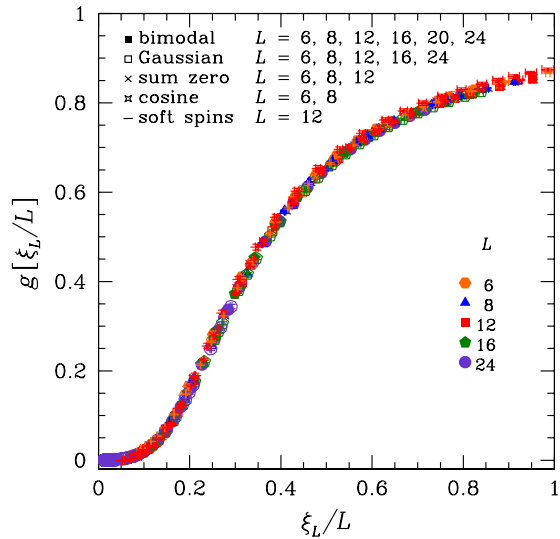}
\vspace*{-1.5cm}

\caption{(Color online)
Binder ratio $g$ as a function of the finite-size correlation length 
$\xi_L/L$ for several system sizes $L \ge 6$ and for all models described in
Sec.~\ref{sec:model}. All data collapse onto a universal curve, thus providing
clear evidence that spin glasses obey universality. Different system sizes are
labeled with different colors, and
different models use different types of symbols as indicated.
}
\label{fig:g_vs_xi_all.eps}
\end{figure}

We have computed two dimensionless quantities, $\xi_L/L$ and $g$, which
intersect at a finite value at the critical temperature. In the previous parts
of this section, we have plotted both
of them against $\beta$. It turns out also
to be useful to plot one of them \textit{against
the other}.\cite{kim:96,joerg:pc}
According to Eq.~(\ref{eq:ci_scale}), $A L^{1/\nu}[\beta - \beta_{\rm
c}] = \widetilde{X}^{-1} (\xi_L / L)$ and so, from Eq.~(\ref{eq:g_scale}), we
can write
\begin{equation}
g =  \widehat{G} \left(\xi_L/L\right) , 
\end{equation}
where $\widehat{G}$ is a universal function. Note that there are \textit{no}
nonuniversal metric factors in this expression. Hence data for
\textit{all} the models
described in Sec.~\ref{sec:model} should collapse when $g$ is plotted against
$\xi_L/L$. This works very well as shown in
Fig.~\ref{fig:g_vs_xi_all.eps} which includes
sizes $L \ge 6$.  Figure \ref{fig:g_vs_xi_all.eps} provides additional
\textit{very strong evidence for universality} in spin glasses.

\section{Summary and Conclusions}
\label{sec:conclusions}

We have studied numerically the phase transition in
a variety of Ising spin-glass models in three dimensions, to
test for universality. Our most detailed simulations are on 
nearest-neighbor models with Gaussian and $\pm J$ interactions, and our
results for them are summarized in Tables \ref{tab:critparams_gauss} and
\ref{tab:critparams_pmj}. A comparison shows that corresponding
estimates for the
exponents $\nu$ and $\eta$ agree well, as do the values of $\xi_L/L$ and $g$
at criticality (labeled $c_0$). This supports universality, as does the plot
of $g$ against $\xi_L/L$, Fig.~\ref{fig:g_vs_xi_all.eps},
where data for \textit{all} the models studied (not
just the Gaussian and $\pm J$ models) collapse onto a single universal curve.

The main unresolved issue is the large difference between the values for $\nu$
obtained from $\xi_L/L$ or $g$ on the one hand and $\chi_{\rm SG}$ on the
other. This is presumably due to systematic errors coming from
corrections to scaling, but unfortunately we
have not been able to incorporate corrections in our analysis since
we do not have data with sufficient precision over a sufficiently large range
of sizes. The errors quoted in this paper are statistical errors only;
systematic errors are not included.
Evidence for strong corrections was found explicitly in
Sec.~\ref{sec:campbell}, where we used
a scaling form for $\chi_{\rm SG}$ proposed
in Ref.~\onlinecite{campbell:06} which differs from the standard form only in
corrections to scaling. From this scaling form, we obtain an estimate for $\nu$ 
from our data for $\chi_{\rm SG}$ which is very different from that obtained
from $\chi_{\rm SG}$ using the standard analysis. This large difference in the
values of $\nu$ from the two methods of analysis does not occur, however,
for our data for $g$ and $\xi_L/L$.

Overall, we have found no evidence for lack of universality, but have found
evidence for strong corrections to scaling. We suspect that the dynamical data
of Refs.~\onlinecite{bernardi:96,mari:99,mari:01,henkel:05,pleimling:05}, which was
interpreted to show lack of universality, more likely shows evidence for
corrections to scaling.

\begin{acknowledgments}
We would like to thank K.~Binder, I.~A.~Campbell, T.~J\"org, K.~Hukushima, 
D.~P.~Landau H.~Takayama, M.~Troyer, and D.~W\"urtz for discussions. 
In particular, we would like to thank T.~J\"org for pointing out the
usefulness of plotting our data in the way shown in
Fig.~\ref{fig:g_vs_xi_all.eps}.
A.P.Y.~acknowledges support from 
the National Science Foundation under NSF Grant No.~DMR 0337049. The 
simulations were performed on the Asgard, Hreidar, and Gonzales clusters
at ETH Z\"urich. We would like to thank K.~Tran for carefully reading the
manuscript.
\end{acknowledgments}

\bibliography{refs,comments}

\begin{thebibliography}{47}
\bibitem[{ple()}]{pleim}
\bibinfo{note}{Reference \onlinecite{henkel:05} claims only that universality
  is violated out of equilibrium.}
\bibitem[{asy()}]{asymmetry}
\bibinfo{note}{Although it is plausible that the data may scale over a wider
  range of $(T-T_{\rm c})$ \textit{above} $T_{\rm c}$ in the analysis of
  Ref.~\onlinecite{campbell:06}, since their method is designed such that the
  high-$T$ behavior is compatible with the scaling function, it is not clear to
  us that the data should also scale over a wider range of $(T-T_{\rm c})$
  \textit{below} $T_{\rm c}$.}
\end{thebibliography}

\end{document}